\documentclass[aps,prl,twocolumn,groupedaddress,showpacs]{revtex4}

\usepackage[dvips]{graphicx,color}

\begin{document}
\title{About the nature of dark matter and dark energy and a model of cosmology that may solve the cosmic coincidence problem}
\author{Isaac Cohen}
\email[]{icohen@brandeis.edu}
\affiliation{Physics Department, Brandeis University, Waltham, MA
02453}

\date{\today}

\begin{abstract}
A model of cosmology, that arises from the  hypothesis that
ordinary matter, dark matter and dark energy are made of the same
stuff, is studied. It is argued that this hypothesis is a
consequence of considering space and time in the same footing. The
model provides a solution to the cosmic coincidence problem
predicting,  without fine-tuning, values of
$\Omega_{m}\approx0.32$, $\Omega_{\Lambda}\approx0.68$ and
$\Omega_{b}\approx0.04$ in agreement with observations. Besides,
the model does not suffer from neither  the flatness nor the
horizon problems,  does not appear to spoil the successes of
cosmic nucleosynthesis calculations  and  allows for the formation
of galaxies and clusters.
\end{abstract}


\maketitle

Current astronomical observations give evidence for a spatially
flat and accelerating universe. Type Ia supernova observations
\cite{rie98} convincingly reject the  deceleration expected from
an Einstein-de Sitter universe and indicate a cosmological
constant of the same order of magnitude as the present mass
density of the universe, anisotropy observations of the Cosmic
Microwave Background radiation (CMB) pointing toward  spatial
flatness \cite{bal00,spe03} . Data from these observations fix
the fraction of matter energy density to critical density to
$\Omega_{m}\sim0.3$ and the corresponding fraction associated to a
cosmological energy to $\Omega_{\Lambda}\sim0.7$. At the same
time, CMB experiments limit the contribution of baryonic matter to
only $\Omega_{b}\sim0.05$, the rest being what is called dark
matter, results  that are in good agreement with the standard Big
Bang nucleosynthesis.

In the same way that Einstein introduced a cosmological constant
and carefully fine-tuned it with the express purpose of contravene
the tendency of matter to produce deceleration, and obtain in that
way  an static universe, now  we must contend with fine tuning
parameters \cite{wei89} such that the present value of the dark
energy (or cosmological constant) be of the same order of
magnitude as the present mass density of the universe. This is the
\textit{cosmic coincidence problem} \cite{note1}.

Alternatives to the cosmological constant has been considered. In
one proposal \cite{rat88}, the dark energy varies with time and is
due to a homogeneous scalar field shifted away from the true
minimum of its potential. These models \cite{zla98} still need a
fine-tuning in order to explain why this quintessence field comes
to dominate only recently \cite{wei00}. The failure to find a
satisfactory model has led to the belief that some sort of
anthropic principle must be at work \cite{gar00}.

Note that the observed dark matter to baryonic matter ratio is
large. Recent measurements of the large-scale distribution of
galaxies gives $\Omega_{m}=0.27\pm0.06$ \cite{ver02}, supporting a
large nonluminous matter to ordinary matter ratio. There have been
proposed many candidates for dark matter, starting from ordinary
material in some nonluminous form, relics of the early universe,
weakly interacting massive particles, supersymmetric particles and
others \cite{tri87}.

It is the purpose of  this note to  propose a model of cosmology
that arises from the hypothesis that ordinary matter, dark matter
and dark energy are made of the same stuff. We will see that such
a model  gives account of the cosmic coincidence problem  in a
natural way. We consider the space time of real events extended to
complex values and allow world lines of particles in some regions
of this extended space time identifying them with either dark
energy or dark matter. The metric in our restricted space time
will be governed by all these contributions to the energy momentum
tensor.

We may argue about the physics behind this extension:
Notwithstanding the fact that in Einstein's special theory of
relativity, space and time appears to be  in the same footing,
there is a basic distinction: Lorentz transformations leave
invariant the form
\begin{eqnarray}
ds^{2}=-dt^{2}+dx^{2}+dy^{2}+dz^{2}, \label{eq1}
\end{eqnarray}
for the interval ds between two events, with a relative minus sign
between the time and space coordinates. We take c=1.  This
relative sign allows the existence of pairs of events with a null
interval and hence to be connected by a light signal, giving then
a distinct physical meaning to the odd coordinate. The distinction
may be hidden if we require, as Minkowski did, that the time
coordinate assume only imaginary values. Symmetry is only obtained
if we allow all coordinates to assume complex values, but now  the
invariant group of transformations becomes the Complex Lorentz
group. This is a well known extension. It is essential in a
rigorous proof of the PCT theorem \cite{str64}. In this framework
there is no preferred direction.

Let us  consider  any event as the origin of space and time
coordinates of a  four dimensional complex space in which the
manifold extending along the real components of the space
directions and the imaginary component of the time direction is
the usual Minkowski space. This manifold contains all the ordinary
events. In this complex space there are, however,  other manifolds
whose points have three real coordinates and one imaginary. There
are  three of such generalized Minkowski spaces, all of them going
through the origin. We may identify each of them by the character
and name of its odd coordinate direction. The null cone that
passes through the origin intersects each one of these spaces
separating the regions in which $ds^{2}>0$ from the ones in which
$ds^{2}<0$. Note that we may think of the odd coordinate direction
as the axis of the projected cone.

If space and time are in the same footing then we should allow for
matter in all these manifolds. Any particle whose  world line is
in any of these  Minkowski spaces should  contribute to the
energy-momentum tensor of matter at the arbitrary point taken as
the origin.

Our basic hypothesis is then that \textit{ordinary matter, dark
matter and dark energy are made of the same stuff. Particles whose
world lines are geodesics inside the null cone in the manifold
along the imaginary direction of t \cite{note2} would be
considered ordinary matter.  Particles whose world lines are
geodesics inside any of the null cones in the manifolds along the
imaginary directions of x, y and z would be considered dark
matter. Particles whose world lines are geodesics outside any of
the null cones of the four manifolds would be considered dark
energy.}

Let us consider any of the  Minkowski spaces along one of the
imaginary space  directions, say the imaginary direction of x.
Inside the null cone, $ds^{2}<0$. In momentum space this means
that the $mass^{2}$ of the corresponding particles  are positive.
We will associate, by hypothesis, these particles to darkmatter
\cite{note3}. For example, particles whose geodesics are along the
imaginary direction of x would contribute to the energy density
and not to the pressure, because these particles will have energy
but no momentum. This can be seen using a complex Lorentz
transformation, the interchange between the x and t coordinates to
connect our frame to the frame in which those particles are at
rest.

Let us consider any of regions  of these four Minskowski spaces
outside the null cone, \textit{i.e,} where $ds^{2}>0$. In momentum
space this means that the $mass^{2}$ of the corresponding
particles  are negative. In a isotropic model, the total
contribution to the energy momentum tensor of such kind of
particles  from these four spaces should behave like an scalar
under Lorentz Transformations. This is consistent with the
hypothesis that associates them to dark energy \cite{gib02}.


For the construction of a model of cosmology we consider an
spatial isotropic and homogeneous universe on our space time
manifold. We assume, as usual,  the validity of Einstein's field
equations. They are
\begin{eqnarray}
R_{\mu\nu}-{1\over 2}g_{\mu\nu}R =-8\pi G~ T_{\mu\nu}, \label{eq2}
\end{eqnarray}
with the sign conventions of \cite{wei72}.  The corresponding
metric has to have the Robertson-Walker form
\begin{eqnarray}
ds^{2}=-dt^{2}+a^{2}(t)\{\frac{dr^{2}}{1-kr^{2}}+r^{2}d\theta^{2}+r^{2}sin^{2}\theta
d\phi^{2}\}, \label{eq3}
\end{eqnarray}
We assume that matter fills homogeneously all the space so that
Einstein's equations reduce to a pair of second order differential
equations on the scale factor a,  and by eliminating $\ddot{a}$,
we obtain Friedmann's equation
\begin{eqnarray}
\dot{a}^{2}+k=\frac{8\pi G}{3}\rho a^{2}. \label{eq4}
\end{eqnarray}
In addition the time dependence of $\varrho$ is given by the
continuity equation,
\begin{eqnarray}
\dot{\rho}=-3\frac{\dot{a}}{a}(\rho+p).\label{eq5}
\end{eqnarray}
This also can be written
\begin{eqnarray}
a\frac{d\rho}{da}=-3(\varrho+p).\label{eq6}
\end{eqnarray}
As it is well known, we need only an equation of state
$p=p(\varrho~)$ to determine $\varrho~$ as a function of the scale
factor a and then, by using equation (\ref{eq4}) to solve for
a(t).

We estimate the equation of state in the following way: The total
density $\rho$ will receive contributions from: i) one share  of
ordinary matter density $\rho_{o}$ coming from geodesics inside
our manifold. ii) Three contributions of darkmatter each one
coming from their  corresponding  manifolds. Each of these
portions will be equal to $\rho_{o}$ because the allowed space for
geodesics inside the regions where $ds^{2}<0$ for all the
manifolds are the same. iii) A portion of dark energy coming from
each of the  four manifolds. The allowed space for geodesics
inside the regions where $ds^{2}>0$ is  three times larger than
inside the regions where $ds^{2}<0$. Then, each of these portions
should be three times bigger than the previous ones. If we
represent the dark energy fluid as made of particles of negative
$mass^{2}$, $m^{2}<0$, the property that they have a momentum that
is invariant under Lorentz Transformations means that these
particles should have an energy density equal to $m/\surd2$. This
gives an additional factor of $1/\surd2$ to the contributions from
dark energy. For a matter-dominated universe we take the ordinary
and dark matter contributions to the pressure as zero and the dark
matter pressure as $p_{\Lambda}=-\rho_{\Lambda}$.  In terms of the
ordinary matter energy density $\rho_{o}$ we have  that the total
density $\rho$ and total pressure p are
\begin{eqnarray}
\rho=\rho_{o}+3\rho_{o}+6\surd 2\rho_{o},~~~~~~~~~~\label{eq7}
p=-6\surd 2\rho_{o}. \label{eq8}
\end{eqnarray}
We eliminate $\rho_{o}$ from the above pair of equations to obtain
the equation of state
\begin{eqnarray}
p=\omega \rho, \label{eq9}
\end{eqnarray}
with $\omega=-(1+\frac{\surd2}{3})^{-1}\approx-.68$.

The equation (\ref{eq4}) for a flat universe, k=0, gives the
critical energy density $\rho_{c}$ as  $\rho_{c}=3H^{2}/8\pi G$,
where H is the Hubble constant. We define a density parameter
$\Omega_{x}$ as the ratio of matter of type x  to the critical
energy density $\rho_{c}$. Equation (\ref{eq7}) gives the relative
proportions of each type of matter. For  matter density
$\rho_{m}$, that  is the sum of  ordinary matter and dark matter
densities, we obtain the value
$\Omega_{m}=1/(1+3/\surd2)\approx0.32$. For dark energy we obtain
$\Omega_{\Lambda}=1/(1+\surd2/3)\approx0.68$. These results are in
agreement with observations \cite{spe03}.

For ordinary matter we obtain $\Omega_{o}=1/(4+6/\surd2)$. If we
take in account that   both particles and antiparticles are
included as ordinary matter  we find that
$\Omega_{b}=1/(8+12/\surd2)\approx0.04$, in agreement with
observations \cite{kra02}.

Solving equation (\ref{eq6}) for $\rho$ subjected to the equation
of state (\ref{eq9}) gives $\rho\propto a^{-3(1+w)}$ and hence we
have that the energy density scales as $\rho\propto
a^{-\frac{6}{2+3\surd2}}$, the exponent being $\approx-.96$.

Using equation~(\ref{eq4}), For the case of a flat universe, we
obtain that the scale factor grows as $a\propto~t^{\alpha}$, with
$\alpha=\frac{2}{3(1+\omega)}$.
$\omega=-(1+\frac{\surd2}{3})^{-1}$ this gives
$\alpha=2/3+\surd2\approx2.08$. If we fix the origin of time such
that the scale factor $a(t)$ is zero at that time, $a(0)=0$, the
present time $t_{0}$ may be identified with the age of the
universe. In that case we obtain a simple relation between the
Hubble constant $H_{0}$ and the age of the universe $t_{0}$,
$H_{0}\equiv
\frac{\dot{a}(t_{0})}{a(t_{0})}=\alpha/t_{0}\approx2.08/t_{0}.$
For the deceleration parameter
$q_{0}\equiv-\ddot{a}(t_{0})\frac{a(t_{0})}{\dot{a}^{2}(t_{0})}$,
we obtain $q_{0}\approx-0.5$.

Taking $H_{0}=13.4~Gyr$ \cite{spe03} one gets that the age of the
universe is $t_{0} \approx28~Gyr$. Recent observational evidence
suggests that at redshift $z\approx6$ is the epoch of reionization
\cite{bec01}, where starlight dominated for the first time in the
history of the universe. Now, the redshift z is given by
\begin{eqnarray}
1+z\equiv\frac{a_{0}}{a(t)}=(\frac{t_{0}}{t})^{\alpha},
\label{eq10}
\end{eqnarray}
For the elapsed time $\Delta t=t_{0}-t$ since an event at time t,
this gives $\Delta t=t_{0}(1-(z+1)^{-1/\alpha})$. The elapsed time
since the the epoch of reionization is then $\approx17~Gyr$. This
is in agreement with studies of stellar ages using radio-isotopes
\cite{sch02}.

If we assume that both ordinary matter and dark matter are
relativistic, {\it i.e.}, a radiation-dominated universe, we have
the same expression for the density as in the equation
(\ref{eq7}), but for the pressure p we have instead that
$p=4/3\rho_{0}-6\surd2\rho_{0}$. The equation of state is like
equation (\ref{eq9}) but now
$\omega_{r}=-\frac{3-\surd2/3}{3+\surd2}\approx-0.57$. The energy
density of radiation scales as $\rho_{r}\propto
a^{-3(1+\omega_{r})}\approx a^{-1.28}$. A flat universe filled
with radiation would grow like $a\propto~t^{\alpha_{r}}$ with
$\alpha_{r}=\frac{1}{2}+\frac{3\surd2}{4}\approx1.56$. Note that
$\alpha/\alpha_{r}=4/3$, the same ratio than in the
Einstein-de~Sitter universe. If one assumes that all energy is in
thermal equilibrium with darkbody radiation, $\rho= \sigma T^{4}$,
we can use the scaling behavior $a\propto~t^{\alpha_{r}}$ for a
flat universe filled with radiation to find the dependence of the
temperature T on time and the parameter $\alpha.$ Equation
(\ref{eq4}) gives then that the temperature T scales like
$T\propto(\alpha_{r}/t)^{1/2}$,
\begin{eqnarray}
T=(\frac{3}{8\pi G \sigma} )^{1/4}(\frac{\alpha_{r}}{t})^{1/2},
\label{eq11}
\end{eqnarray}
where ${\sigma}$ is the radiation density constant.

If the universe is flat, ($\Omega=1$), then it will remain so for
all time. We  may rewrite equation (\ref{eq4}) as
$|1-\Omega|=1/\dot{a}^{2}$. This shows that in a nearly flat
universe $|1-\Omega|\propto~t^{-2.16}$ for a matter-dominated
universe or $\propto~t^{-1.12}$ for a radiation-dominated
universe. We don't have to fine-tune the initial conditions in
order that the universe looks flat now. In consequence, the model
we are considering does not have the flatness problem.

The particle horizon \cite{rin56} defined  by
$r_{hor}(t_{0})\equiv~\int_{0}^{t_{0}}\frac{dt}{a}$ gives $\infty$
for a flat universe. This means that the range of causal
interactions is infinite. There is no horizon problem
\cite{wei72,mis73} in our model. Note that the satisfactory
solution for the flatness and horizon problems that we have found
are due to the fact that the model share with inflationary models
\cite{gut81} the property that the scale factor is accelerating
$\ddot{a}>0$. The similarities between inflationary models and the
present model ends here because in inflationary models there is
first an era of inflation pumped by an scalar field followed by a
normal era in which the scale factor is decelerating. Here the
scale factor keeps accelerating.

It has been argued \cite{fre87} that   comparable amounts of  dark
energy and matter in the  early history of the universe would
spoil the successes of cosmic nucleosynthesis calculations. We
will show that this is not the case here. The argument relies in a
calculation showing that the ratio $\eta$ of the number of baryons
to the number of photons exhibits a strong dependence on the
temperature if there is a substantial component of dark energy and
the observation that the temperature at which nucleosynthesis
takes place depends on  this ratio \cite{fre87}. To be able to do
the calculation they assume thermal equilibrium and that dark
matter exchanges energy either with radiation or with matter.
Here, we need only to require thermal equilibrium. We have that to
maintain a Planck distribution, the energy per particle must
redshift like the temperature, $T\propto 1/\surd t
\propto~a^{-3(1+\omega)}$. Taking in account the scaling behavior
of $\rho_{r}$ we find that the photon number density scales as
$n_{\gamma}\propto a^{-\frac{3}{2\alpha_{r}}}$. On the other hand,
the scaling behavior of the number density of baryons may be
written as $n_{b}=a^{-2/\alpha}$. Their scaling behavior is, then,
the same and in consequence the ratio $\eta\equiv
n_{b}/n_{\gamma}$ is independent of the temperature.

It is important to see whether the model allows for the  the
formation of galaxies and clusters. We may wonder if a density
perturbation may grow without limitations. It has been found that
a cosmological constant  should  not be comparable to matter
density at the time of galaxy formation \cite{wei87}. In the model
we are considering, the dark energy is not constant but is
decreasing.  It is easy to see that in this case the proof does
not give any bound. But as long as we take the scaling properties
for the matter density already obtained, there is no bound even if
for the sake of the argument we take the dark energy
$\rho_{\Lambda}$ as a constant. Let us follow Weinberg
\cite{wei87} modelling a perturbation as a sphere within which
there is a uniform excess density $\Delta\rho(t)$, its evolution
been controlled  by a Friedmann equation
\begin{eqnarray}
\dot{a}^{2}+\Delta k=\frac{8\pi
G}{3}(\rho_{m}+\Delta\rho_{m}+\rho_{\Lambda}) a^{2}. \label{eq12}
\end{eqnarray}
If we take in account that the density scales as $\rho\propto
a^{-3(1+\omega)}$ we have that
\begin{eqnarray}
a^{3(1+\omega)}(\varrho+\Delta\rho)=const. \label{eq13}
\end{eqnarray}
The problem is whether the perturbed scale factor $a(t)$ increases
to a maximum value and then collapse or whether there is no bound.
The critical point occurs when the right hand side of equation
(\ref{eq12}) is equal to $\Delta k$. It is obtained that there is
no minimum if $\omega<-1/3$. This is the case here.  The
conclusion is, then, that in this model a perturbation can
increase without limit.



\begin{thebibliography}{99}
\bibitem{note2}
We use the same letters x, y, z and  t to indicate the real space
and  time coordinates and the corresponding complex ones. No
confusion should arise because the distinction should be clear
from the context.
\end{thebibliography}

\end{document}